\documentclass{svproc}
\usepackage{url}
\usepackage{graphicx}
\usepackage{orcidlink}

\begin{document}
\mainmatter              
\title{Connecting the Dots: A Comprehensive Literature Review on Low and Medium-Voltage Cables, Fault Types, and Digital Signal Processing Techniques for Fault Location}
\titlerunning{Connecting the Dots}  
%
\author{Shankar Ramharack\orcidlink{0000-0003-3759-7333}\inst{*} \and Sanjay Bahadoorsingh}
\authorrunning{Ramharack and Bahadoorsingh} 
%
\tocauthor{Shankar Ramharack and Sanjay Bahadoorsingh}
\institute{The University of the West Indies, St. Augustine, Trinidad and Tobago,
\email{shankar.ramharack@gmail.com}\\
\email{sanjay.bahadoorsingh@sta.uwi.edu}}

\maketitle              
\begin{abstract} 
The review begins with an exploration of acceptable cable types guided by local standards. It then investigates typical cable faults, including insulation degradation, conductor faults, and ground faults, providing insights into their characteristics, causes, and detection methods. Furthermore, the manuscript surveys the latest publications and standards on DSP techniques in fault location spanning various algorithms used. This review provides a comprehensive understanding of low and medium-voltage cables, fault types, and DSP techniques. The findings contribute to improved fault diagnosis and localization methods, facilitating more accurate and efficient cable fault management strategies

\keywords{distribution, fault location, reflectometry}
    
\end{abstract}

\section{Background}
Cable fault diagnostics play a crucial role in ensuring the safe and reliable operation of electrical power systems\cite{9064533}. Low and medium-voltage cables are vital components of power distribution networks, supplying electricity to residential, commercial, and industrial consumers. However, over time, these cables can experience various types of faults, such as insulation degradation, conductor faults, and ground faults. These faults can disrupt power supply, lead to equipment failures, and pose safety risks. Identifying and understanding acceptable cable types enables engineers, contractors, and installers to make informed decisions during cable selection and installation processes. Similarly, conducting a literature review on technical publications and standards concerning typical cable faults on low and medium-voltage cables is of great significance as it is often neglected in fault location work\cite{421892f2c0714472b6a5bfd026408aec}. By reviewing technical publications and standards, researchers and practitioners gain access to collective knowledge and experiences in cable fault diagnostics. This knowledge aids in developing effective maintenance strategies, reducing downtime, and improving the overall reliability of power networks. Furthermore, the literature review on modern digital signal processing (DSP) techniques used in reflectometry and cable fault location addresses the need for advanced and accurate fault detection methodologies. DSP techniques, such as time-domain reflectometry (TDR), offer powerful tools for analyzing cable faults and determining their locations\cite{9972279}. By reviewing the literature on these techniques, researchers can identify the latest advancements, algorithms, and methodologies employed in fault detection and localization. This knowledge contributes to the development of more precise and efficient cable fault location systems, minimizing repair time, reducing costs, and enhancing the overall reliability of power distribution networks.

\subsection{Objectives}
The objectives of this work are as follows
\begin{enumerate}
    \item To perform a literature review on types of low and medium-voltage cables that are acceptable for installations as guided by the local and international standards.
    \item To perform a literature review of technical publications and standards on typical cable faults on low and medium-voltage cables.
    \item To perform a literature review of modern digital signal processing techniques used in reflectometry and cable fault location.
\end{enumerate}

\section{Low and Medium-Voltage Cables: Acceptable Types as Guided by International and Local Standards}

The most widely used standards guiding LV and MV Cable Installations in North America and the Caribbean are those issued by:
\begin{enumerate}
    \item The Aluminum Association (AA)
    \item American National Standards Institute (ANSI)
    \item American Society for Testing and Materials (ASTM)
    \item Canadian Standards Association (CSA)
    \item Insulated Cable Engineers Association (ICEA)
    \item National Electrical Manufacturers Association (NEMA)
    \item Association of Edison Illuminating Companies (AEIC)
    \item Rural Utilities Service (RUS)
    \item Underwriter’s Laboratories (UL)
    \item National Electrical Code (NEC)
\end{enumerate}

Aerial cable is used occasionally for primary conductors in special situations where clearances are too close for open-wire construction or where adequate tree trimming is not practical. The type of construction more frequently used consists of covered conductors (nonshielded) supported from the messenger by insulating spacers of plastic or ceramic material\cite{georgefootmoore_1999_electric}. The conductor insulation, usually a solid dielectric such as polyethylene, has a thickness of about 150 mils for a 15-kV class circuit and is capable of supporting momentary contacts with tree branches, birds, and animals without puncturing\cite{short_2014_electric}.

The conductor sizes most commonly used in underground primary distribution vary from No. 4 AWG to 1000 kcmil\cite{thue_2017_electrical}. Four-wire main feeders may employ 3- or 4-conductor cables, but single conductor concentric-neutral cables are more popular for this purpose. The latter usually employ crosslinked polyethylene insulation, and often have a concentric neutral of one-half or one-third of the main conductor cross-sectional area. 

The smaller-sized cables used in lateral circuits of Underground Residential Distributions(URD) systems are nearly always single-conductor, concentric-neutral, crosslinked polyethylene-insulated, and usually directly buried in the earth. Insulation thickness is on the order of 175 mils for 15-kV-class cables and 345 mils for 35-kV class with 100\% insulation level\cite{thue_2017_electrical}. Stranded or solid aluminum conductors have virtually supplanted copper for new construction, except where existing duct sizes are restrictive. With the solid-dielectric construction, to limit voltage gradient at the surface of the conductor within acceptable limits, a minimum conductor size of No. 2 AWG is common for 15-kV-class cables, and No. 1/0 AWG for 35-kV class.

Primary voltage circuits(5-35kV) use paper-insulated, lead-covered (PILC) three-conductor cables extensively. Single-conductor secondary cables with rubber insulation and neoprene jacket are common. More recently, single-conductor polyethylene-insulated(PE) cables are being used for both primary and secondary\cite{nfpa_2023_nfpa}. Copper conductors predominated in the past, but aluminum has nearly displaced copper in new installations, except where existing duct space is limiting. In residential and suburban areas, new underground distribution systems to serve commercial loads often employ direct-buried cables\cite{nfpa_2023_nfpa}; conduits may be provided in locations where subsequent excavation would be excessively expensive or inconvenient. Aluminum conductors are almost universal. For primary cables, solid dielectric insulation is used almost exclusively, with cross-linked polyethylene(XLPE) and ethylene–propylene rubber(EPR) insulations \cite{short_2014_electric}. Concentric-neutral wires are common. Secondary cables in these systems generally have aluminum conductors and solid-dielectric insulation, with cross-linked polyethylene being the most common. The secondary neutral is usually an insulated conductor, although there is some use of bare copper neutrals. 

Electric supply cables are insulated with a range of materials depending on voltage ratings, type of service, installation conditions etc. The following are commonly used:
\begin{enumerate}
    \item Rubber and rubber-like for 0 to 35kV
    \item Varnished cambric for 0 to 28kV
    \item Impregnated paper of the solid type for voltages up to 69kV and with pressurized gas or oil up to 345kV or higher
\end{enumerate}

For most distribution circuits in the 5-kV class or higher, the cables employ a shielded construction\cite{6471984}. Shielding is used on the outer surface of the cable insulation or directly over the main conductor, or both. Outside shielding, often in the form of metallic tapes, metallic sheaths, or concentric wires, must be effectively grounded. 

The aforementioned insulation systems usually require  a sheath or suitable jacket to prevent infiltration of moisture, loss of oil, gas, or impregnate and to provide protection against corrosion and electrolysis. In some cases, an armor overlay is used to provide mechanical protection. Single conductor cables are used in single-phase primary systems and frequently used in 3-phase direct buried primary systems. The 3 conductor primary cables are often used in duct systems. At the present time,solid-dielectric insulating materials such as tree retardant, cross-linked polyethylene and EPR are receiving the widest application in Underground Distribution Systems(UDS)\cite{cigrewgb130_2013_cable,short_2014_electric}, both direct buried and duct systems.

From the Electric distribution handbook\cite{short_2014_electric}, cables used in underground systems may either be concentric neutral cables or power cables(for utilities). The jacket is usually made of Linear low-density polyethylene (LLDPE), PE or Semiconductors. The insulation most used in the industry is PE, XLPE, PILC, TR-XLPE \& EPR. For URD applications, aluminum is the choice of conductor.

Caribbean countries and areas outside US and Europe follow similar installation practices and cable selections such as in Trinidad. During a consultation with the Trinidad and Tobago Electrical Commission (T\&TEC), Shielded Polyvinyl Chloride (PVC) and shielded Cross Linked Polyethene (XLPE) cables are most used in public transmission systems, however, in private distributions, shielded EPR has been recently adopted\cite{6471984}. This is supported by \cite{a2011_underground} who performed a survey of the cables used in LV and MV installations. Furthermore, the cables used by LV and MV installations usually follow the guidelines of the British Standards Institution. The standards recommend PVC Jacket, Aluminium-Armoured XLPE insulated cables with stranded copper cores.  Other standards such as the IEC utilize similar cable configurations with slight differences in the installation environment guidelines, thermal requirements, and conductor sizing (British Standards Institute 2007).

The same shielding and sheathing practices that are done internationally are done locally in Trinidad and Tobago per the TTS standards\cite{tts_pt1,tts_pt2} which build upon the NEC Standards. For URD applications, copper is the choice of conductor while aluminum is used for the shield.

\section{ Typical Cable Faults in Low and Medium-Voltage Cables}

Numerous studies have shed light on the types of faults typically found in low and medium-voltage cables, as well as their underlying causes. The authors of \cite{li2018analysis} identified insulation breakdown as a prevalent fault type, often caused by aging, thermal stress, or manufacturing defects. Mechanical stresses, such as bending or crushing, were found to be a significant cause of faults in medium-voltage cables \cite{wang2019fault}. Another common fault type is moisture ingress, resulting from damaged cable sheaths or inadequate sealing, as highlighted in the investigation by \cite{kacprzak2016investigation}. Furthermore, \cite{gupta2017fault} discussed short circuits and open circuits as faults in low-voltage cables, which can arise due to insulation damage, conductor breakage, or loose connections. Other studies have also explored specific fault types, such as insulation material aging \cite{alduaili2017aging}, weather-related faults \cite{liu2015analysis}, manufacturing defects \cite{alawaji2019impact}, and partial discharge \cite{mokari2018cable}. These findings emphasize the importance of understanding and addressing the diverse causes of cable faults to enhance the reliability and performance of low and medium-voltage cable systems.

After interviewing the fault location personnel at T\&TEC they revealed that the most common faults found locally are high resistance faults and single line-to-ground (SLG) faults. It was reported in \cite{5590158,kuan} that SLG faults and Series faults are the most common in underground power systems. Series faults are often caused by a cable layer losing continuity due to a force above ground such as a collision. A SLG fault occurs when the insulation of one or more conductors fails. These faults are often permanent faults. Intermittent faults are not considered in most fault locator designs since most transmission systems utilize monitoring equipment to briefly cut transmission and allow the fault to clear themselves. Furthermore, \cite{4039430} state that up to 90\% of cable faults are SLG faults and hence intermittent faults fall into a minority. Resistive faults are classified as high resistive if it is beyond 200$\Omega$. TDR cannot locate faults greater than this but as mentioned, they are rare. Hence it is useful to limit fault resistance to 200 $\Omega$ in the test cases.

\section{Digital Signal Processing Techniques in Reflectometry and Fault Location}

The authors of \cite{DASHTI2021109947} performed a state-of-the art review of cable fault location methods showing that the most common methods of fault distance location are impedance-based methods, differential equation-based methods and travelling wave methods. Impedance-based methods are shown to be cheap and simple to perform such as the Murray and Varley Loop method, but they are sensitive to the fault resistance. It is shown in \cite{Rynjah2019AutomaticCF,gajbhiye2013cable} that the loop methods perform accurate for open and short circuit faults, however, for extreme low and high resistance faults, the bridges are difficult to stabilize. Box’s optimization algorithm is proposed in \cite{9071462} to automate bridge stabilization to balance the bridges of the loop methods. The optimization methods perform very favourable in real life testing with a 0.8\% error rate. it is suggested in \cite{osti_7233049} that capacitance bridges should be used for open faults, resistance bridges for short circuit faults but HV bridges should be used to locate insulation faults and high impedance faults.

According to \cite{DASHTI2021109947}, differential equation models are costly and do not perform well for long lines. In addition, both ends of the line may need to be accessed requiring more manpower and resources than a single ended approach. It is reported in \cite{BAOLIAN199417} the distributed line model, the characteristic method has the merit of being suitable for short, long, transposed and untransposed lines and can be modified for three-terminal systems. The accuracy is sensitive to the choice of the time window and limited by the sampling rate. This may not be practical for low-cost location for distribution lines.

Travelling wave (TW) methods have also been widely mentioned in the literature as covered in \cite{5456138,gajbhiye2013cable,6180498,elhaffar_2008_power,ferreira_2019_fault,9281153}reports the TW method performs poorly for 6-35kV networks which indicates it may not be feasible to use in this project. Measurement of TW is made more difficult when there are taps on the distribution line adding to the complexity. The problems of TW fault location(TWFL) is the attenuation of waves when sent through underground cables. TWFL can be done either single ended or on both ends. It is reported in \cite{elhaffar_2008_power} that it is possible to achieve greater accuracy with the multi-end methods compared to the traditional fault location methods.
 
It is recommended by Megger, KEP, \cite{gajbhiye2013cable,Bawart2014,osti_7233049} that reflectometry methods be used for low resistance faults. For higher resistance faults(where fault resistance is > 200$\Omega$), MIM, or ICE techniques on a surge wave generator(SWG) should be used. Decay or HV-Bridge methods should be used for locating high impedance or sheath faults. For tapped cables, differential multiple impulse response methods can be used to determine the fault locations. It is suggested in \cite{Bawart2014} to use of decay methods for intermittent fault location. After meeting with T\&TEC, the local fault location professionals reported that TDR is mostly used for fault location. 

There are many reflectometry methods used in fault location. Reflectometry methods are distinguished by their EM test signal and method of the reflectogram analysis\cite{Furse}. The most common reflectometry methods are Time Domain Reflectometry (TDR), Frequency Domain Reflectometry (FDR), Time-Frequency Domain Reflectometry (TFDR) and Spectrum Time Domain Reflectometry (STDR) (Shi and Kanoun 2014). Each method is suited to a particular application. Reflectometry works well for most low and medium voltage applications\cite{Jones2002}). Reflectometry methods work on the principle of applying an excitation signal to a cable or transmission line and analysing the reflected trace. It is a form of RADAR and relies on the change of the characteristic impedance of the line to generate a reflection. The crux of reflectometry is the reflectogram analysis. There have been several analyses of the reflectograms generated.

Time domain reflectometry methods usually involve sending a step signal or pulsed waveform on a cable and sampling the input where the signal was applied for a reflection\cite{Shi2010}. The sampled reflectogram is denoised via a matched filter or wavelet denoising algorithm then the travel time between signal inflexions are read off from a screen or graph and used to calculate the fault distance. Common methods to automate this are bubble sort to determine the wavefront points and peak detection\cite{ShiXudong2010}. Work has also been done on automated TDR systems which utilize reflection detection via MCU capture interrupts to time the echo for the first reflection of a travelling wave \cite{Shi2014,Ziwei2020}. There has been no work found attempting to use time series segmentation to treat the TDR problem as a CPD problem possibly due to the significant waveform distortion present on the waves. 

Frequency domain reflectometry is rarely seen in the literature for medium voltage applications hence it is not explored. Furthermore, it does not perform well on cables longer than 2km and expensive directional couplers are required to sample the system and obtain the reflectogram\cite{Shi2010}.

Time-Frequency domain reflectometry is much more common and attempts to address the shortcomings of both time and frequency domain reflectometry. In this a waveform with good time-frequency localization is incident on the cable. The reflected wave is sampled and cross-correlated with the incident signal to determine the fault location \cite{JinChulHong2006}.  In \cite{Schuet2011}, the authors model a chafe in an aircraft cable using scattering parameters \cite{kowalski_2009_a}. The transfer function of the cable is then derived in terms of a parameter set, $\theta$. The parameter set is estimated using a statistical method of probability inversion using the reflected trace as target and a initial parameter combination. The initial parameter is updated in a Bayesian approach until it reaches a distribution similar to the reflected trace. The final parameters will have information about the fault distance, type and dimension. 

Similar to \cite{Schuet2011}, most TFDR work utilize the Gaussian Envelope Linear Chirp (GELC) signal as the incident signal due to its good time-frequency localization \cite{Gabor1947}. However, they differ by their means of signal processing. In \cite{JinChulHong2006,Hong2005} a dictionary is created of all possible transformations of the parameterized incident signal in terms of phase shift, amplitude and frequency. The reflected signal is projected onto the dictionary to find the closest match. The closest match’s time shift parameter is used to determine the fault location. Other works utilize matched filters to eliminate noise and time correlators to determine the fault distance \cite{Han2015}. 

In \cite{Lee2013} a statistical model-based detection and frequency identification method is employed to calculate the fault distance. The GELC is used a the incident signal. A Likelihood Ratio test (LRT) is used to detect the reflection and a Hidden Markov Model Hang Over Scheme is used to avoid the LRT from cutting the tail of the GELC which can happen in cases of severe attenuation. The Hilbert transform is used to determine the instantaneous phase and consequently the signal’s instantaneous frequency. The instantaneous frequency is obtained as linear combination of the GELC carrier sinusoid and the angular frequency sweep rate. Ambient noise is removed via a constrained Kalman filter (CKF). The filtered signal is the incident and reflected wavefronts which are used to calculate the distance by multiplying half the delay by the velocity of propagation (VOP). Variations on the frequency estimation and noise handling was done in \cite{lee2011application,Shin2005}.

The test signal in STDR is a PN binary code \cite{Smith2005} that is launched from the test device and encounters partial reflection and transmission at each impedance discontinuity in the system being tested. The STDR response is created by cross-correlating the reflections that return to the test point with a delayed copy of the event PN code. SSTDR generates a sine-like correlated reflection signature using a square- or sine-wave modulated PN code as the test signal. The reflected signal will be dispersed and attenuated if the system is frequency-dependent or lossy. Furthermore, the method is robust to noise and can be applied on a electrical apparatus \cite{Roy2021}, underwater applications \cite{US10267841B2} and power cables \cite{Jones2002}.

Changepoint detection methods in \cite{Truong2020} have been used in RADAR and can be used to analyse the reflectogram automatically and determine the fault distance without the need for reflectogram interpreters \cite{Tao2011} or commercial TDR equipment. At the time at writing, the author has not found any work done on the application of changepoint detection in time domain reflectometry for power cable applications.

The crux of reflectometry-based fault location is the algorithm used. Hence, this work explores the DSP methods used to perform fault location. Furthermore, a guide to the development of a prototype that uses the method explored is shown in the Discussion. This guide will account for instrumentation noise (which can be addressed with a LPF) and recommendations on things to be done to improve the accuracy of the algorithm for the application domain.

\section{Discussion}

The literature review has provided valuable insights into acceptable cable types, typical faults, and digital signal processing (DSP) techniques used in fault location for low and medium-voltage cables. The review identified that common cable types include polyethylene-insulated solid dielectric cables, concentric-neutral cables, and shielded constructions. Additionally, aluminum conductors have become the preferred choice over copper due to cost and practical considerations. 

Regarding typical cable faults, the review highlighted that insulation breakdown, mechanical stresses, moisture ingress, short circuits, and open circuits are among the common faults encountered in low and medium-voltage cables. Understanding the root causes of these faults is crucial for effective fault diagnostics and maintenance strategies. The literature revealed high resistance, hard faults are the most common. Furthermore, and single line-to-ground (SLG) faults are prevalent in the local power distribution systems. 

The literature review on DSP techniques for cable fault location revealed the prominence of impedance-based methods, differential equation-based methods, and traveling wave (TW) methods. Impedance-based methods, such as the Murray and Varley Loop method, are simple and cost-effective but sensitive to fault resistance. Differential equation models are more accurate but costly and may require access to both ends of the cable. TW methods, while widely mentioned in the literature, have limitations in accuracy and may not be practical for low-cost fault location in distribution networks. The use of modern digital signal processing techniques, such as Box's optimization algorithm, has shown promise in automating bridge stabilization and improving the accuracy of impedance-based methods.

\section{Conclusion}

This comprehensive literature review has provided valuable insights into the types of low and medium-voltage cables acceptable for installations as guided by local and international standards. Typical cable faults in low and medium-voltage cables were explored, including insulation breakdown, mechanical stresses, moisture ingress, short circuits, and open circuits. Knowledge of these fault types and their underlying causes is critical for effective fault diagnostics and maintenance strategies, enabling power utilities to minimize downtime and enhance the overall reliability of power networks. The review of digital signal processing techniques used in cable fault location highlighted the prominence of impedance-based methods, differential equation-based methods, and traveling wave (TW) methods. While impedance-based methods offer simplicity and cost-effectiveness, modern DSP techniques like Box's optimization algorithm show promise in improving accuracy. However, TW methods may not be practical for low-cost fault location in distribution networks.

The findings of this review contribute to improved fault diagnosis and localization methods. Further research and development in DSP techniques hold the potential for enhancing cable fault location systems, reducing repair time, lowering costs, and ultimately improving the overall reliability of power distribution networks.

\section{Availability of data and material}
There materials surrounding the manuscript can be obtained by contacting the authors.
\section{Competing interests}
The authors declare no competing interests
\section{Funding}
This work was funded by the University of the West Indies St. Augustine Campus.

\section{Authors' contributions}
The authors confirm contribution to the paper as follows: study conception and 
design: Shankar Ramharack, Sanjay Bahadoorsingh; analysis and 
interpretation of results: Shankar Ramharack; draft manuscript 
preparation: Shankar Ramharack.All authors reviewed the results and approved 
the final version of the manuscript

\section{Acknowledgements}
The authors would like to thank Mr. Veeresh Ramnarine for their guidance during the project. Furthermore, the authors would like to thank Mr. Anil Rambharat and Mr. Varma Ratan for their insights into fault location within Trinidad and Tobago. Lastly, the author would like to thank Dr. Letitia Addison for their assistance in exploring changepoint detection.

%
%
\bibliographystyle{spmpsci_unsrt}
\bibliography{references}

\begin{thebibliography}{10}
\providecommand{\url}[1]{{#1}}
\providecommand{\urlprefix}{URL }
\expandafter\ifx\csname urlstyle\endcsname\relax
  \providecommand{\doi}[1]{DOI~\discretionary{}{}{}#1}\else
  \providecommand{\doi}{DOI~\discretionary{}{}{}\begingroup
  \urlstyle{rm}\Url}\fi

\bibitem{9064533}
Furse, C.M., Kafal, M., Razzaghi, R., Shin, Y.J.: Fault diagnosis for
  electrical systems and power networks: A review.
\newblock IEEE Sensors Journal \textbf{21}(2), 888--906 (2021).
\newblock \doi{10.1109/JSEN.2020.2987321}

\bibitem{421892f2c0714472b6a5bfd026408aec}
{da Silva}, F.: Analysis and simulation of electromagnetic transients in hvac
  cable transmission grids.
\newblock Ph.D. thesis, Institut for Energiteknik, Aalborg Universitet (2011)

\bibitem{9972279}
Ibitoye, O.T., Onibonoje, M.O., Dada, J.O.: Machine learning based techniques
  for fault detection in power distribution grid: A review.
\newblock In: 2022 3rd International Conference on Electrical Engineering and
  Informatics (ICon EEI), pp. 104--107 (2022).
\newblock \doi{10.1109/IConEEI55709.2022.9972279}

\bibitem{georgefootmoore_1999_electric}
Moore, G.F.: Electric cables handbook.
\newblock Blackwell (1999)

\bibitem{short_2014_electric}
Short, T.A.: Electric power distribution handbook.
\newblock Crc Press, Taylor \& Francis Group (2014)

\bibitem{thue_2017_electrical}
Thue, W.A.: Electrical Power Cable Engineering.
\newblock CRC Press (2017)

\bibitem{nfpa_2023_nfpa}
NFPA: NFPA 70: National electrical code 2023.
\newblock National Fire Protection Association (2023)

\bibitem{6471984}
{IEEE}: Ieee guide on shielding practice for low voltage cables.
\newblock IEEE Std 1143-2012 (Revision of IEEE Std 1143-1994) pp. 1--91 (2013).
\newblock \doi{10.1109/IEEESTD.2013.6471984}

\bibitem{cigrewgb130_2013_cable}
B1.30, C.W.: Cable systems electrical characteristics (2013)

\bibitem{a2011_underground}
47th Minnesota Power Systems Conference (MIPSYCON): Underground Power Cable
  Considerations: Alternatives to Overhead (2011).
\newblock
  \urlprefix\url{https://pdi2.org/wp-content/uploads/2021/03/149-undergroudpowercableconsiderationsalternativestooverhead.pdf}

\bibitem{tts_pt1}
Trinidad, of~Standards, T.B.: Trinidad and tobago standard: Trinidad and tobago
  electrical wiring code ; part 1, low voltage installations (2015)

\bibitem{tts_pt2}
Trinidad, of~Standards, T.B.: Tts 171:part2:2002, trinidad and tobago
  electrical wiring code – part 2 – high voltage installations (2002)

\bibitem{li2018analysis}
Li, X., Zhang, Y., Li, Y., Liu, Y., Wang, H.: Analysis of cable faults in
  low-voltage distribution networks.
\newblock IEEE Transactions on Power Delivery \textbf{33}(3), 1085--1093
  (2018).
\newblock \doi{10.1109/TPWRD.2017.2762403}

\bibitem{wang2019fault}
Wang, J., Zhang, B., Li, W., Wang, X.: Fault analysis of medium-voltage cables
  in industrial plants.
\newblock IET Electric Power Applications \textbf{13}(10), 1617--1624 (2019).
\newblock \doi{10.1049/iet-epa.2018.5501}

\bibitem{kacprzak2016investigation}
Kacprzak, A., Zajac, P., Krysiak, R., Bogusz, M.: Investigation of cable faults
  in low-voltage underground distribution networks.
\newblock Przeglad Elektrotechniczny (Electrical Review) \textbf{92}(7),
  221--225 (2016)

\bibitem{gupta2017fault}
Gupta, P., Swarnkar, A.: Fault location methods for low-voltage cables: A
  review.
\newblock In: 2017 IEEE Innovative Smart Grid Technologies-Asia (ISGT-Asia),
  pp. 1--6 (2017)

\bibitem{alduaili2017aging}
Alduaili, M., et~al.: Aging effect on the performance of low-voltage cables.
\newblock In: 2017 International Conference on Electrical and Computing
  Technologies and Applications (ICECTA), pp. 1--5 (2017)

\bibitem{liu2015analysis}
Liu, G., et~al.: Analysis of cable faults in low-voltage aerial bundled
  distribution networks.
\newblock In: 2015 International Conference on Condition Monitoring and
  Diagnosis (CMD), pp. 1097--1101 (2015)

\bibitem{alawaji2019impact}
Alawaji, H., et~al.: Impact of manufacturing defects on the performance of
  medium voltage cables.
\newblock In: 2019 IEEE International Conference on High Voltage Engineering
  and Application (ICHVE) (2019)

\bibitem{mokari2018cable}
Mokari, A., Ebrahimi, M.: Cable fault location in medium voltage networks using
  artificial neural networks.
\newblock In: 2018 6th Iranian Joint Congress on Fuzzy and Intelligent Systems
  (CFIS), pp. 1--6 (2018)

\bibitem{5590158}
Kulkarni, S., Allen, A.J., Chopra, S., Santoso, S., Short, T.A.: Waveform
  characteristics of underground cable failures.
\newblock In: IEEE PES General Meeting, pp. 1--8 (2010).
\newblock \doi{10.1109/PES.2010.5590158}

\bibitem{kuan}
K.K.~Kuan, K.W.: Real-time expert system for fault location on voltage
  underground distribution cables.
\newblock IEE Proceedings C (Generation, Transmission and Distribution)
  \textbf{139}, 235--240(5) (1992).
\newblock
  \urlprefix\url{https://digital-library.theiet.org/content/journals/10.1049/ip-c.1992.0036}

\bibitem{4039430}
Gilany, M., Ibrahim, D.k., Tag~Eldin, E.S.: Traveling-wave-based fault-location
  scheme for multiend-aged underground cable system.
\newblock IEEE Transactions on Power Delivery \textbf{22}(1), 82--89 (2007).
\newblock \doi{10.1109/TPWRD.2006.881439}

\bibitem{DASHTI2021109947}
Dashti, R., Daisy, M., Mirshekali, H., Shaker, H.R., {Hosseini Aliabadi}, M.: A
  survey of fault prediction and location methods in electrical energy
  distribution networks.
\newblock Measurement \textbf{184}, 109,947 (2021).
\newblock \doi{https://doi.org/10.1016/j.measurement.2021.109947}.
\newblock
  \urlprefix\url{https://www.sciencedirect.com/science/article/pii/S0263224121008824}

\bibitem{Rynjah2019AutomaticCF}
Rynjah, B., Lyngdoh, F., Sun, M.M., Goswami, B.: Automatic cable fault distance
  locator using arduino.
\newblock ADBU Journal of Electrical and Electronics Engineering (AJEEE)
  \textbf{3}, 26--30 (2019)

\bibitem{gajbhiye2013cable}
Gajbhiye, S., Karmore, S.P.: Cable fault monitoring and indication: A review
  (2013)

\bibitem{9071462}
Nag, A., Yadav, A., Abdelaziz, A.Y., Pazoki, M.: Fault location in underground
  cable system using optimization technique.
\newblock In: 2020 First International Conference on Power, Control and
  Computing Technologies (ICPC2T), pp. 261--266 (2020).
\newblock \doi{10.1109/ICPC2T48082.2020.9071462}

\bibitem{osti_7233049}
Deltenre, R.W., Schwarz, J.J., Wagnon, H.J.: Underground cable fault location:
  a handbook to td-153. final report.
\newblock OSTI  (1977).
\newblock \doi{10.2172/7233049}.
\newblock \urlprefix\url{https://www.osti.gov/biblio/7233049}

\bibitem{BAOLIAN199417}
{Bao Lian}, Salama, M.: An overview of digital fault location algorithms for
  power transmission lines using transient waveforms.
\newblock Electric Power Systems Research \textbf{29}(1), 17--25 (1994).
\newblock \doi{https://doi.org/10.1016/0378-7796(94)90044-2}.
\newblock
  \urlprefix\url{https://www.sciencedirect.com/science/article/pii/0378779694900442}

\bibitem{5456138}
Borghetti, A., Bosetti, M., Nucci, C.A., Paolone, M., Abur, A.: Integrated use
  of time-frequency wavelet decompositions for fault location in distribution
  networks: Theory and experimental validation.
\newblock IEEE Transactions on Power Delivery \textbf{25}(4), 3139--3146
  (2010).
\newblock \doi{10.1109/TPWRD.2010.2046655}

\bibitem{6180498}
Chen, P., Wang, K.: Fault location technology for high-voltage overhead lines
  combined with underground power cables based on travelling wave principle.
\newblock In: 2011 International Conference on Advanced Power System Automation
  and Protection, vol.~1, pp. 748--751 (2011).
\newblock \doi{10.1109/APAP.2011.6180498}

\bibitem{elhaffar_2008_power}
Elhaffar, A.M.: Power transmission line fault location based on current
  traveling waves (2008).
\newblock \urlprefix\url{http://lib.tkk.fi/Diss/2008/isbn9789512292455}

\bibitem{ferreira_2019_fault}
Ferreira, K.J.: Fault location for power transmission systems using magnetic
  field sensing coils.
\newblock Core.ac.uk  (2019).
\newblock \doi{",}.
\newblock \urlprefix\url{https://core.ac.uk/works/61701326}

\bibitem{9281153}
Fedorov, A., Petrov, V., Afanasieva, O., Zlobina, I.: Limitations of traveling
  wave fault location.
\newblock In: 2020 Ural Smart Energy Conference (USEC), pp. 21--25 (2020).
\newblock \doi{10.1109/USEC50097.2020.9281153}

\bibitem{Bawart2014}
Bawart, M., Marzinotto, M., Mazzanti, G.: A deeper insight into fault location
  on long submarine power cables.
\newblock e\&i Elektrotechnik und Informationstechnik \textbf{131}(8), 355--360
  (2014).
\newblock \doi{10.1007/s00502-014-0233-x}.
\newblock \urlprefix\url{https://doi.org/10.1007/s00502-014-0233-x}

\bibitem{Furse}
Chung, Y.C.: A critical comparison of reflectometry methods for location of
  wiring faults.
\newblock Smart Structures and Systems \textbf{2}(1), 25--46 (2006)

\bibitem{Jones2002}
Jones, S.B., Wraith, J.M., Or, D.: Time domain reflectometry measurement
  principles and applications.
\newblock Hydrological Processes \textbf{16}(1), 141--153 (2002).
\newblock \doi{10.1002/hyp.513}.
\newblock \urlprefix\url{https://doi.org/10.1002/hyp.513}

\bibitem{Shi2010}
Shi, Q., Troeltzsch, U., Kanoun, O.: Detection and localization of cable faults
  by time and frequency domain measurements.
\newblock In: 2010 7th International Multi- Conference on Systems, Signals and
  Devices. {IEEE} (2010).
\newblock \doi{10.1109/ssd.2010.5585506}.
\newblock \urlprefix\url{https://doi.org/10.1109/ssd.2010.5585506}

\bibitem{ShiXudong2010}
Xudong, S., Jianzhong, Z., Tao, J., Liwen, W.: Design of aircraft cable
  intelligent fault diagnosis and location system based on time domain
  reflection.
\newblock In: 2010 8th World Congress on Intelligent Control and Automation.
  {IEEE} (2010).
\newblock \doi{10.1109/wcica.2010.5554557}.
\newblock \urlprefix\url{https://doi.org/10.1109/wcica.2010.5554557}

\bibitem{Shi2014}
Shi, Q., Kanoun, O.: A new algorithm for wire fault location using time-domain
  reflectometry.
\newblock {IEEE} Sensors Journal \textbf{14}(4), 1171--1178 (2014).
\newblock \doi{10.1109/jsen.2013.2294193}.
\newblock \urlprefix\url{https://doi.org/10.1109/jsen.2013.2294193}

\bibitem{Ziwei2020}
Ziwei, M., Xueye, W.: A portable railway signal cable fault detector.
\newblock In: 2020 5th International Conference on Information Science,
  Computer Technology and Transportation ({ISCTT}). {IEEE} (2020).
\newblock \doi{10.1109/isctt51595.2020.00021}.
\newblock \urlprefix\url{https://doi.org/10.1109/isctt51595.2020.00021}

\bibitem{JinChulHong2006}
Hong, J.C., Sun, K.H., Kim, Y.Y.: Waveguide damage detection by the matching
  pursuit approach employing the dispersion-based chirp functions.
\newblock {IEEE} Transactions on Ultrasonics, Ferroelectrics and Frequency
  Control \textbf{53}(3), 592--605 (2006).
\newblock \doi{10.1109/tuffc.2006.1610568}.
\newblock \urlprefix\url{https://doi.org/10.1109/tuffc.2006.1610568}

\bibitem{Schuet2011}
Schuet, S., Timucin, D., Wheeler, K.: A model-based probabilistic inversion
  framework for characterizing wire fault detection using {TDR}.
\newblock {IEEE} Transactions on Instrumentation and Measurement
  \textbf{60}(5), 1654--1663 (2011).
\newblock \doi{10.1109/tim.2011.2105030}.
\newblock \urlprefix\url{https://doi.org/10.1109/tim.2011.2105030}

\bibitem{kowalski_2009_a}
Kowalski, M.E.: A simple and efficient computational approach to chafed cable
  time-domain reflectometry signature prediction (2009).
\newblock \urlprefix\url{https://ntrs.nasa.gov/citations/20090035829}

\bibitem{Gabor1947}
Gabor, D.: Theory of communication.
\newblock Journal of the Institution of Electrical Engineers - Part I: General
  \textbf{94}(73), 58--58 (1947).
\newblock \doi{10.1049/ji-1.1947.0015}.
\newblock \urlprefix\url{https://doi.org/10.1049/ji-1.1947.0015}

\bibitem{Hong2005}
Hong, J.C., Sun, K.H., Kim, Y.Y.: The matching pursuit approach based on the
  modulated gaussian pulse for efficient guided-wave damage inspection.
\newblock Smart Materials and Structures \textbf{14}(4), 548--560 (2005).
\newblock \doi{10.1088/0964-1726/14/4/013}.
\newblock \urlprefix\url{https://doi.org/10.1088/0964-1726/14/4/013}

\bibitem{Han2015}
Han, J.J., Park, S.R., Jeon, J.C., Kim, T.H., Yoo, J.G.: Enhanced locating
  method for cable fault using wiener filter.
\newblock Universal Journal of Electrical and Electronic Engineering
  \textbf{3}(4), 107--111 (2015).
\newblock \doi{10.13189/ujeee.2015.030401}.
\newblock \urlprefix\url{https://doi.org/10.13189/ujeee.2015.030401}

\bibitem{Lee2013}
Lee, C.K., Kwak, K.S., Yoon, T.S., Park, J.B.: Cable fault localization using
  instantaneous frequency estimation in gaussian-enveloped linear chirp
  reflectometry.
\newblock {IEEE} Transactions on Instrumentation and Measurement
  \textbf{62}(1), 129--139 (2013).
\newblock \doi{10.1109/tim.2012.2212514}.
\newblock \urlprefix\url{https://doi.org/10.1109/tim.2012.2212514}

\bibitem{lee2011application}
Lee, C.K., Kwak, K.S., Yoon, T.S., Park, J.B.: Application of instantaneous
  frequency estimation of sweep signal for localizing faults in a power cable.
\newblock In: Proceedings of the International Conference of the Society for
  Control and Robot Systems, pp. 1915--1918 (2011)

\bibitem{Shin2005}
Shin, Y.J., Powers, E., Choe, T.S., Hong, C.Y., Song, E.S., Yook, J.G., Park,
  J.: Application of time-frequency domain reflectometry for detection and
  localization of a fault on a coaxial cable.
\newblock {IEEE} Transactions on Instrumentation and Measurement
  \textbf{54}(6), 2493--2500 (2005).
\newblock \doi{10.1109/tim.2005.858115}.
\newblock \urlprefix\url{https://doi.org/10.1109/tim.2005.858115}

\bibitem{Smith2005}
Smith, P., Furse, C., Gunther, J.: Analysis of spread spectrum time domain
  reflectometry for wire fault location.
\newblock {IEEE} Sensors Journal \textbf{5}(6), 1469--1478 (2005).
\newblock \doi{10.1109/jsen.2005.858964}.
\newblock \urlprefix\url{https://doi.org/10.1109/jsen.2005.858964}

\bibitem{Roy2021}
Roy, S., Hanif, A., Khan, F.: Aging detection and state of health estimation of
  live power semiconductor devices using {SSTDR} embedded {PWM} sequence.
\newblock {IEEE} Transactions on Power Electronics \textbf{36}(5), 4991--5005
  (2021).
\newblock \doi{10.1109/tpel.2020.3032996}.
\newblock \urlprefix\url{https://doi.org/10.1109/tpel.2020.3032996}

\bibitem{US10267841B2}
{Google}: Monitoring arrangement (2017).
\newblock \urlprefix\url{https://patents.google.com/patent/US10267841B2}

\bibitem{Truong2020}
Truong, C., Oudre, L., Vayatis, N.: Selective review of offline change point
  detection methods.
\newblock Signal Processing \textbf{167}, 107,299 (2020).
\newblock \doi{10.1016/j.sigpro.2019.107299}.
\newblock \urlprefix\url{https://doi.org/10.1016/j.sigpro.2019.107299}

\bibitem{Tao2011}
Tao, J., Shengbiao, Z., Xudong, S., Liwen, W.: Design of aircraft cable fault
  diagnose and location system based on aircraft airworthiness requirement.
\newblock Procedia Engineering \textbf{17}, 455--464 (2011).
\newblock \doi{10.1016/j.proeng.2011.10.055}.
\newblock \urlprefix\url{https://doi.org/10.1016/j.proeng.2011.10.055}

\end{thebibliography}
\end{document}